\documentclass[12pt]{article}
\usepackage{amsmath}
\usepackage{amssymb}
\usepackage{euscript}

\usepackage[hang]{subfigure}

\usepackage{epsfig}

\usepackage{cite}

\usepackage{alltt}

\begin{document}

\begin{titlepage}
\begin{flushright}
\bf IFJPAN-IV-2013-10
\end{flushright}
\vspace*{1 cm}

\begin{center}
{\Large\bf 
 Markov Chain Mote Carlo solution of BK equation
 \\ \vspace{2mm}
  through Newton--Kantorovich method 
}
\vskip 1cm

{\large 
  Krzysztof Bo\.zek$^{a,b}$, 
  Krzysztof Kutak$^a$
  and
   Wies\l{}aw P\l{}aczek$^b$
}

\vskip 5mm
{\em
   $^a$Institute of Nuclear Physics, Polish Academy of Sciences,\\
  ul.\ Radzikowskiego 152, 31-342 Krakow, Poland.
  \\ \vspace{2mm}
 $^b$Marian Smoluchowski Institute of Physics, Jagiellonian University,\\
  ul. Reymonta 4, 30-059 Krakow, Poland.
  }
\end{center}


\vskip 1.5cm
\begin{abstract}
\noindent
We propose a new method for Monte Carlo solution of  non-linear integral equations by combining the 
Newton--Kantorovich method for solving non-linear equations with the Markov Chain Monte Carlo (MCMC) method for solving 
linear equations. 
The Newton--Kantorovich method allows to express the non-linear equation as a system of the linear equations which then 
can be treated by the MCMC (random walk) algorithm. We apply this method to the Balitsky--Kovchegov (BK) equation
describing evolution of gluon density  at low $x$. Results of numerical computations show that the MCMC method
is both precise and efficient. The presented algorithm may be particularly suited for solving more complicated and higher-dimensional
non-linear integral equation, for which traditional methods become unfeasible.

\begin{flushleft}
\end{flushleft}
\end{abstract}

\vskip 3cm

\begin{flushleft}
\bf IFJPAN-IV-2013-10
\end{flushleft}

\end{titlepage}

\section{Introduction}
The Large Hadron Collider (LHC) at CERN provides an opportunity to scan parton densities in the proton 
over a wide domain of parton kinematics. 
This allows for detailed studies of dynamical effects taking place during evolutions of partons. An example of the dynamical phenomena 
which is particularly interesting in hadronic processes is saturation of gluon density \cite{Gribov:1984tu}.
At high energies the dominant contribution to evolution of system of partons comes from splittings of gluons and this leads to rapid growth of  the gluon density and, as a consequence, to fast rise of corresponding cross sections. 
The unitarity constraints suggest that eventually the growth of the gluon density should slow down due to possible effects of fusion of the gluons, leading to its saturation. 
And indeed, there is a growing evidence that the saturation occurs in high-energy hadron collision processes \cite{Albacete:2010pg,Dumitru:2010iy,Kutak:2012rf,Dusling:2012cg}. 

The physics of the saturation at an inclusive level is described within the perturbative QCD by
\cite{JalilianMarian:1997jx,JalilianMarian:1997gr,Kovner:2000pt,Iancu:2000hn,Balitsky:1995ub,Kovchegov:1999yj},
 and at an exclusive level by the equations proposed in \cite{Kutak:2011fu,Kutak:2012yr,Kutak:2012qk}.   
The standard approach in search of the saturation with the JIMWLK/BK evolution equation is to solve the equation that provides the gluon density, and then to convolute the solutions with appropriate matrix elements which specify the final states.
This approach has some limitations, since it does not allow for the full simulation of a scattering event, as is for example modelled by Monte Carlo event generators \cite{Sjostrand:2006za,Bahr:2008pv,Jung:2010si,Kharraziha:1997dn,Andersen:2011zd}. The Monte Carlo event generators allow for exact treatment of kinematical effects, storing information on emitted partons, etc.
The particularly useful method of performing Monte Carlo simulation is based on a Markov Chain (random walk) approach 
\cite{Forsythe:1950}.
In this approach,  the evolution process occurs over evolution `time' which could be, for example, an energy scale. Such an evolution can be interpreted as a Markovian probabilistic process. 
The main advantage of this approach is that one performs the full Monte Carlo simulation in a forward process, without the need to pretabulate the solution of the considered equation. 
In the so-called backward evolution method, first the appropriate equation is solved and the corresponding parton density is pretabulated,
and then the  actual Monte Carlo evolution (random walk) is performed backward in a `time' variable to simulate the scattering process with a probability distribution given by the respective parton density.
The application of the Markov Chain Monte Carlo (MCMC) algorithm is, however, not straightforward for equations which model the saturation effects, since it works only for the linear evolution equations. To our best knowledge, such an algorithm has not been, so far, applied to the non-linear evolution equations.

In the present paper we develop a method which allows to perform the MCMC-based solution of the non-linear equation%
\footnote{For another approach to modelling of non-linear effects with Monte Carlo techniques 
we refer the reader to \cite{Flensburg:2011kk}.}. 
We apply this method to the BK equation.
The key idea is to apply a well-convergent method for solving the non-linear integral equation  and to combine it with a Monte Carlo algorithm designed for solving the linear integral equations.
We have found that particularly well-suited for this purpose is the Newton--Kantorovich method \cite{polyanin}. 
It relies on representing the non-linear integral equation as a system of the linear equations (see eq.~(\ref{eq:BK-NK1-intro})--(\ref{eq:BK-NK3-intro})), to which the MCMC algorithm can be applied. The whole procedure can be done in iterative manner and it does not require one to provide the solution of the considered equation in advance. 
Furthermore, it can be used for solving the exclusive saturation equations, as proposed in \cite{Kutak:2011fu,Kutak:2012yr,Kutak:2012qk}, and even more complicated and higher-dimensional non-linear integral equations, for which other numerical methods
are unfeasible (inefficient, ustable, etc.).  
This might also be a first step in constructing the Monte Carlo event generator for modelling the saturation effects in the fully exclusive way.

The paper is organized as follows. In Section~2 we introduce the Newton--Kantorovich method for the BK equation. In Section~3 we describe the Markov Chain Monte Carlo algorithm. In Section 4 we combine the Newton--Kantorovich method with the Monte Carlo algorithm to provide the solution of the BK equation. We also compare our solution with the one provided by the {\sf BKSolver} package \cite{Enberg:2005cb}. 

\section{Newton--Kantorovich method for BK equation}
\label{sec:BK}

Let us consider the leading-order in $\alpha_s\ln(1/x)$ Balitsky--Kovchegov (BK) equation for the Weiz\"acker--Williams gluon density:
\begin{equation}
\begin{aligned}
\Phi(x,k^2) = \Phi^0(x,k^2) & +\bar{\alpha}_s 
\int_{x}^{1} \frac{dz}{z} \int_{0}^{\infty}\frac{d l^2}{l^2}\left[\frac{l^2\Phi(x/z,l^2)-k^2\Phi(x/z,k^2)}{|k^2-l^2|}+\frac{k^2 \Phi(x/z,k^2)}{\sqrt{4l^4+k^4}}\right]
\\
&-\frac{\bar{\alpha}_s}{\pi R^2}\int_{x}^{1}\frac{d z}{z}\,\Phi^2(x/z,k^2),
\end{aligned}
\label{eq:BK1}
\end{equation}
where $\Phi^0(x,k^2)$ is a driving term,
 $\bar{\alpha}_s = (N_c\alpha_s)/\pi$ (in our calculations we use $\alpha_s=0.2$), $k\equiv k_{\perp}$ is the transverse gluon momentum, $x$ is the fraction of the longitudinal proton momentum
 carried by the gluon, and hereinafter we set $R=1/\sqrt{\pi}$.
First, we perform the following change of variables: $y = -\ln x,\; t=y + \ln z \Rightarrow x = \mathrm{e}^{-y},\; x/z = \mathrm{e}^{-t}$ and simplify the notation by skipping exponents of arguments of the above functions, i.e. $\Phi(\mathrm{e}^{-y},\ldots) \rightarrow \Phi(y,\ldots)$, etc., to obtain:
\begin{equation}
\begin{aligned}
\Phi(y,k^2)= \Phi^0(y,k^2)&+\bar{\alpha}_s  \int_{0}^{y} dt \int_{0}^{\infty}\frac{d l^2}{l^2}\left[\frac{l^2\Phi(t,l^2)-k^2\Phi(t,k^2)}{|k^2-l^2|}+\frac{k^2 \Phi(t,k^2)}{\sqrt{4l^4+k^4}}\right]
\\
&-\bar{\alpha}_s \int_{0}^{y}dt\,\Phi^2(t,k^2).
\end{aligned}
\label{eq:BK2}
\end{equation} 
 Introducing a dimensionful constant $\mu^2$, the dimensional integration variable $l^2$ can be replaced by  $\lambda=\ln(l^2/\mu^2)$, for which we have $d\lambda=dl^2/l^2$. Introducing also $\kappa=\ln(k^2/\mu^2)$, we get
\begin{equation}
\begin{aligned}
\Phi(y,\kappa)= \Phi^{0}(y,\kappa)&+\bar{\alpha}_s  \int_{0}^{y} dt \int_{0}^{\infty} d\lambda\left[\frac{e^{\lambda}\Phi(t,\lambda)-e^{\kappa}\Phi(t,{\kappa})}{|e^{\kappa}-e^{\lambda}|}+\frac{e^{\kappa} \Phi(t,{\kappa})}{\sqrt{4e^{2\lambda}+e^{2\kappa}}}\right]
\\
&-\bar{\alpha}_s \int_{0}^{y}dt\,\Phi^2(t,{\kappa}),
\end{aligned}
\label{eq:BK3}  
\end{equation}
where the notation is again simplified: we use $\kappa$ instead of $k^2$ in arguments of $\Phi$ and $\Phi^0$, and drop the dependence on the scale $\mu^2$, which is obviously hidden in both functions. 
This is the two-dimesional non-linear integral equation of the form:
\begin{equation}
\Phi(y,\kappa)= \Phi^{0}(y,\kappa)+ \int_{0}^{y} dt \int_{0}^{\infty} d\lambda\, K\left(y,t,\kappa,\lambda,\Phi(t,\lambda)\right),
\label{eq:NLeq}
\end{equation}
 with the kernel
\begin{equation}
\begin{aligned}
K\left(y,t,\kappa,\lambda,\Phi(t,\lambda)\right) & = \bar{\alpha}_s\left[\frac{e^{\lambda}\Phi(t,{\lambda})-e^{\kappa}\Phi(t,{\kappa})}{|e^{\kappa}-e^{\lambda}|}+\frac{e^{\kappa} \Phi(t,{\kappa})}{\sqrt{4e^{2\lambda}+e^{2\kappa}}}\right]
\\ &
-\bar{\alpha}_s\, \delta(\lambda-\kappa)\, \Phi^2(t,{\lambda})
\label{eq:BKkern}
\end{aligned}
\end{equation}

One can linearize the above equation by expanding the kernel $K\left(y,t,\kappa,\lambda,\Phi(t,\lambda)\right)$ in the Taylor series with
respect to $\Phi(t,\lambda)$ about some $\bar{\Phi}(t,\lambda)$ and retaining only the first two terms:
\begin{equation}
\begin{aligned}
K\left(y,t,\kappa,\lambda,\Phi(t,\lambda)\right) &= K\left(y,t,\kappa,\lambda,\bar{\Phi}(t,\lambda)\right)+K^{'}_{\Phi}\left(y,t,\kappa,\lambda,\bar{\Phi}(t,\lambda)\right)\left[\Phi(t,\lambda)-\bar{\Phi}(t,\lambda)\right]\\
&+\mathcal{O}\left(\left[\Phi(t,\lambda)-\bar{\Phi}(t,\lambda)\right]^2\right).
\end{aligned}
\label{eq:TaylorSeries}
\end{equation}
                                              
Introducing the function $\Psi(t,\lambda)=\Phi(t,\lambda)-\bar{\Phi}(t,\lambda)$ and assuming $|\Psi(t,\lambda)|\ll 1$, we can replace 
eq.~(\ref{eq:BK3}) with the following set of equations:
\begin{eqnarray}
\Phi(y,\kappa) \hspace{-2mm} &=& \hspace{-2mm} \bar{\Phi}(y,\kappa) + \Psi(y,\kappa),
\label{eq:BK-NK1-intro}  \\
\Psi(y,\kappa) \hspace{-2mm} &=& \hspace{-2mm}
\Lambda(y,\kappa) + \int_{y_0}^y dt \int_{0}^{+\infty} d\lambda \,K^{'}_{\Phi}\left(y,t,\kappa,\lambda,\bar{\Phi}(t,\lambda)\right)
\Psi(t,\lambda),
\label{eq:BK-NK2-intro}  \\
 \Lambda(y,\kappa) \hspace{-2mm} &=& \hspace{-2mm}
 \Phi^0(y,\kappa) + \int_{y_0}^y dt \int_{0}^{+\infty}  d\lambda \, K\left(y,t,\kappa,\lambda,\bar{\Phi}(t,\lambda)\right) - \bar{\Phi}(y,\kappa),
\label{eq:BK-NK3-intro}
\end{eqnarray}
where the partial derivative $K^{'}_{\Phi}\left(y,t,\kappa,\lambda,\bar{\Phi}(t,\lambda)\right)$ of the kernel $K$ with respect to 
$\Phi(t,\lambda)$ reads
\begin{equation}
\begin{aligned}
K^{'}_{\Phi}\left(y,t,\kappa,\lambda,\bar{\Phi}(t,\lambda)\right) &= \bar{\alpha}_s\left[\frac{e^{\lambda}\delta(\lambda-{\kappa})-e^{\kappa}}{|e^{\kappa}-e^{\lambda}|}+\frac{e^{\kappa}}{\sqrt{4e^{2\lambda}+e^{2\kappa}}}\right]
\\&
-2\bar{\alpha}_s\, \delta(\lambda-\kappa)\, \bar{\Phi}(t,{\lambda}).
\end{aligned}
\label{eq:BKkernp}
\end{equation}

The above set of equations can be solved by iteration, which leads to the Newton--Kantorovich form of the BK equation:
\begin{eqnarray}
\Phi_n(y,\kappa) \hspace{-2mm} &=& \hspace{-2mm} \Phi_{n-1}(y,\kappa) + \Psi_{n-1}(y,\kappa),
\label{eq:BK-NK1}  \\
\Psi_{n-1}(y,\kappa) \hspace{-2mm} &=& \hspace{-2mm}
\Lambda_{n-1}(y,\kappa) + \int_{y_0}^y dt \int_{0}^{+\infty} d\lambda \,K^{'}_{\Phi}\left(y,t,\kappa,\lambda,\Phi_{n-1}(t,\lambda)\right)
\Psi_{n-1}(t,\lambda),
\label{eq:BK-NK2}  \\
 \Lambda_{n-1}(y,\kappa) \hspace{-2mm} &=& \hspace{-2mm}
 \Phi^0(y,\kappa) + \int_{y_0}^y dt \int_{0}^{+\infty}  d\lambda \, K\left(y,t,\kappa,\lambda,\Phi_{n-1}(t,\lambda)\right) - \Phi_{n-1}(y,\kappa).
\label{eq:BK-NK3}  
\end{eqnarray}
As one can see, instead of the single non-linear integral equation (\ref{eq:NLeq}) we have now the iterative series of the linear 
integral equations (\ref{eq:BK-NK2}),  associated with the auxiliary integrals of eq.~(\ref{eq:BK-NK3}).
This can be solved by the standard iteration (successive approximation) method.
The two-dimesional integrations can be performed directly with the standard numerical quadratures or, alternatively, one may expand the integrands in series of the Chebyshev polynomials, at least in one integration variable. 
The main advantage of the above decomposition is that the integral equation (\ref{eq:BK-NK2}) is linear, and thus one can try to solve it 
by using the MCMC algorithm.

\section{MCMC method}
\label{sec:MCsolBK}

Our goal in this section is to construct a MCMC solution of eq.~(\ref{eq:BK-NK2}) 
which is the Volterra--Fredholm linear integral equation of the second kind.
We can write immediately its iterative solution:
\begin{equation}
\begin{aligned}
&\Psi_{n-1}(y,\kappa)  =  \Lambda_{n-1}(y,\kappa)\\ 
&+ \sum_{m=1}^{\infty} \prod_{i=1}^m\left[ \int_{y_0}^y dt_i\int_{0}^{+\infty} d\lambda_i\, \theta(t_{i-1}-t_i)\,K^{'}_{\Phi}\left(t_{i-1},t_{i},\lambda_{i-1},\lambda_i,\Phi_{n-1}(t_i,\lambda_i)\right)\right]\Psi_{n-1}(t_m,\lambda_m).
\end{aligned}
\label{eq:VFeit-sol}
\end{equation}
Since the integration limits do not depend on the variable $\kappa$, there is no ordering in the integration variable $\lambda$ 
and at any step it
can take an arbitrary value. Due to the ordering in the integration variable $t$, it will play a role of the evolution time in the corresponding
MCMC algorithm. We propose the following MCMC algorithm:
\begin{enumerate}
\item
Start a random walk (Markov chain) from the point $(t_0,\lambda_0)=(y,\kappa)$.
\item
Being at the point $(t_i,\lambda_i)$:
\begin{itemize}
\item[(i)]
 generate a random step in the $t$-direction $\tau_{i+1}=t_{i+1}-t_i<0$ according to some probability density function (pdf)
$\rho(\tau)$, with the normalisation contidion 
$$
\int_{-\infty}^0d\tau\,\rho(\tau)= 1;
$$
\item[(ii)]
for a given value $\tau_{i+1}$, generate a random step in the $\lambda$ direction: $\xi_{i+1}=\lambda_{i+1}-\lambda_i$ according to 
some pdf $\eta_{\tau_{i+1}}(\xi)$, with the normalisation condition
$$
\int_{0}^{+\infty}d\xi\,\eta_{\tau_{i+1}}(\xi)= 1,
$$
where $\eta_{\tau}(\xi)$ is the pdf of the variable $\xi$ depending on the parameter $\tau$ (if it does not depend on
this parameter, then $\xi$ can be generated completely independently of $\tau$).
\end{itemize}
Both $\rho(\tau_{i+1})$ and $\eta_{\tau_{i+1}}(\xi_{i+1})$ may, in general, depend also on  $t_i$ and $\lambda_i$, 
i.e. the distribution of the step size $(\tau_{i+1},\xi_{i+1})$ may differ from step to step. 
\item
Stop the random walk when some  $t_{m+1}$ jumps beyond the lower $t$-integral limit, 
i.e. $t_{m+1} \leq y_0$ following the  sequence $t_0 > t_1 > t_2 > \ldots > t_m > y_0$.
\item
To each trajectory
\begin{equation}
\gamma_m=\left\{(t_0,\lambda_0), (t_1,\lambda_1),\ldots,(t_m,\lambda_m)\colon y=t_0 > t_1 > t_2 > \ldots > t_m > y_0 \geq t_{m+1}\right\}
\label{eq:VF-traject}
\end{equation} 
assign the von Neumann--Ulam weight%
\footnote{Originally, a similar weight was proposed by J.\ von Neumann and S.\ Ulam for matrix inversion.}%
 \cite{Forsythe:1950}:
\begin{equation}
\begin{aligned}
&w(y,\kappa) = \frac{v(\gamma_n)\Lambda_{n-1}(t_m,\lambda_m)}{R(t_m)},\\
&v(\gamma_i)=\frac{\,K^{'}_{\Phi}\left(t_{i-1},t_{i},\lambda_{i-1},\lambda_i,\Phi_{n-1}(t_i,\lambda_i)\right)}{\rho(\tau_i)\,\eta_{\tau_i}(\xi_i)}\,v(\gamma_{i-1}),\quad v(\gamma_0)=1,
\end{aligned}
\label{eq:weights}
\end{equation} 
where 
\begin{equation}
R(t) = \int_{-\infty}^{y_0 - t} d\tau\,\rho(\tau)
\label{eq:vNUgam0}
\end{equation} 
is the probalility of a single jump beyond $y_0$ from the point $t$.\\
Instead of the von Neumann--Ulam weight one may use the Wasow weight%
\footnote{This kind of weight was originally proposed by W.\ Wasow to improve efficiency of the von Neumann--Ulam method for matrix inversion.}%
 \cite{Wasow:1952}:
\begin{equation}
w(y,\kappa) = \sum_{i=0}^m v(\gamma_i)\,\Lambda_{n-1}(t_i,\lambda_i).
\label{eq:VF-WasowRV}
\end{equation}
\item
Repeat the above steps $N$ times and compute the MCMC estimate of $\Psi_{n-1}(y,\kappa)$ as well as its statistical error (standard deviation): 
\begin{equation}
\begin{aligned}
\hat{\Psi}_{n-1}(y,\kappa) & = \frac{1}{N}\sum_{k=1}^N w_k(y,\kappa), \\
\hat{\sigma}_{\hat{\Psi}_{n-1}}(y,\kappa) & = 
\frac{1}{\sqrt{N-1}} \, \sqrt{\frac{1}{N} \sum_{k=1}^N w_k^2(y,\kappa) - \left[\hat{\Psi}_{n-1}(y,\kappa)\right]^2}.
\end{aligned}
\label{eq:VF-estMC}
\end{equation} 
\end{enumerate}
where $w_k(y,\kappa)$ is the trajectory weight (of eq.~(\ref{eq:weights}) or eq.~(\ref{eq:VF-WasowRV})) computed in the $k$th 
repetition of the above steps 1--4.

One can prove that expectation values of the weights $w(y,\kappa)$ of eqs.~(\ref{eq:weights}) and (\ref{eq:VF-WasowRV}) 
satisfy the equation (\ref{eq:BK-NK2}). 
The most straightforward way to do this is to first obtain general expressions for contributions to the weights 
coming from the trajectory of the length $m$, and then, based on that, construct the corresponding expectation values. 

\section{Numerical results}
\label{sec:nuerics}
In this section we present an implementation of the MCMC algorithm. We perform computations on a 2-dimensional lattice of points -- in the rapidity $y$ and in the dimensionless variable $\kappa$, corresponding to the transverse momentum $k_{\perp}$. The results presented here correspond to the lattice with $100$ points in $y$ distributed linearly from $0.0$ to $8.1$,  and $128$ points in $\kappa$ linearly spread in the range $[0.0, 10.6]$. The $k_{\perp}$ dimension is introduced to the problem through the constant $\mu^2$ which shows up in the driving term $\Phi^0(y,\kappa)$:
\begin{equation}
\Phi^0(y,\kappa)=\exp\left(-\frac{\mu^2e^{\kappa}}{\rm GeV^2}\right).
\label{eq:DrivingTerm}
\end{equation}
In our computations we have used $\mu^2=5\cdot 10^{-3}\,$GeV$^2$.

In eq.~(\ref{eq:BK3}) one can see that the integration over $\lambda$ goes to infinity. In order to perform numerical calculations we need to introduce a certain cut-off. 
The driving term of eq.~(\ref{eq:DrivingTerm}) as well as the solution of the BK equation vanish for large $k_{\perp}$, 
thus introducing the upper cut-off on $\kappa$ does not affect the solution considerably. 

For the pdfs $\rho(\tau)$ and $\eta(\xi)$ we use the exponential distributions:
\begin{equation}
\rho(\tau_i)=e^{\tau_i}, \qquad \eta_{\lambda_{i-1}}(\xi_i)=e^{-(\xi_i+\lambda_{i-1})}=e^{-\lambda_{i}},
\label{eq:PDFs}
\end{equation}
and thus the random variables $\tau_i$ and $\lambda_i$ can be generated as follows:
\begin{equation}
\tau_i=\ln U_i, \quad \lambda_i=-\ln V_i, 
\label{eq:random-variables}
\end{equation}
where $U_i$ and $V_i$ are the random variables uniformly distributed between $0$ and $1$, i.e. $U_i,\,V_i\in U(0,1)$.
This choice does its job in the case of the above BK equation, however one can improve the convergence of the MCMC method by using 
the pdfs that are better adjusted to the problem. 
Ideally, the product of these pdfs should be as close as possible to the kernel $K^{'}_{\Phi}$, so that 
all the weights $v$ in eq.~(\ref{eq:weights}) be close to $1$. 
In fact our choice of the pdfs seems to be good enough as we have reached a sufficient precision generating only $1000$ trajectories  
for each iteration of eq.~(\ref{eq:BK-NK2}). 
The results presented here correspond to $15$ iterations of the set of equations (\ref{eq:BK-NK1})--(\ref{eq:BK-NK3}).
Without special optimisations it took only about $20$ minutes of CPU time to generate all the results on a 2.2~GHz Intel Pentium Dual-Core 
processor with the GNU/Linux operating system using only one CPU core.

As stated in the previous section, one can use either the von Neumann--Ulam or the Wasow weights in the MCMC procedure. Our implementation of the MCMC algorithm has been tested with both of them, giving the same results 
(differences not visible in the plots like the one presented below). 

\begin{figure}[ht] 
\centering
\subfigure[]{\includegraphics[width=0.49\textwidth]{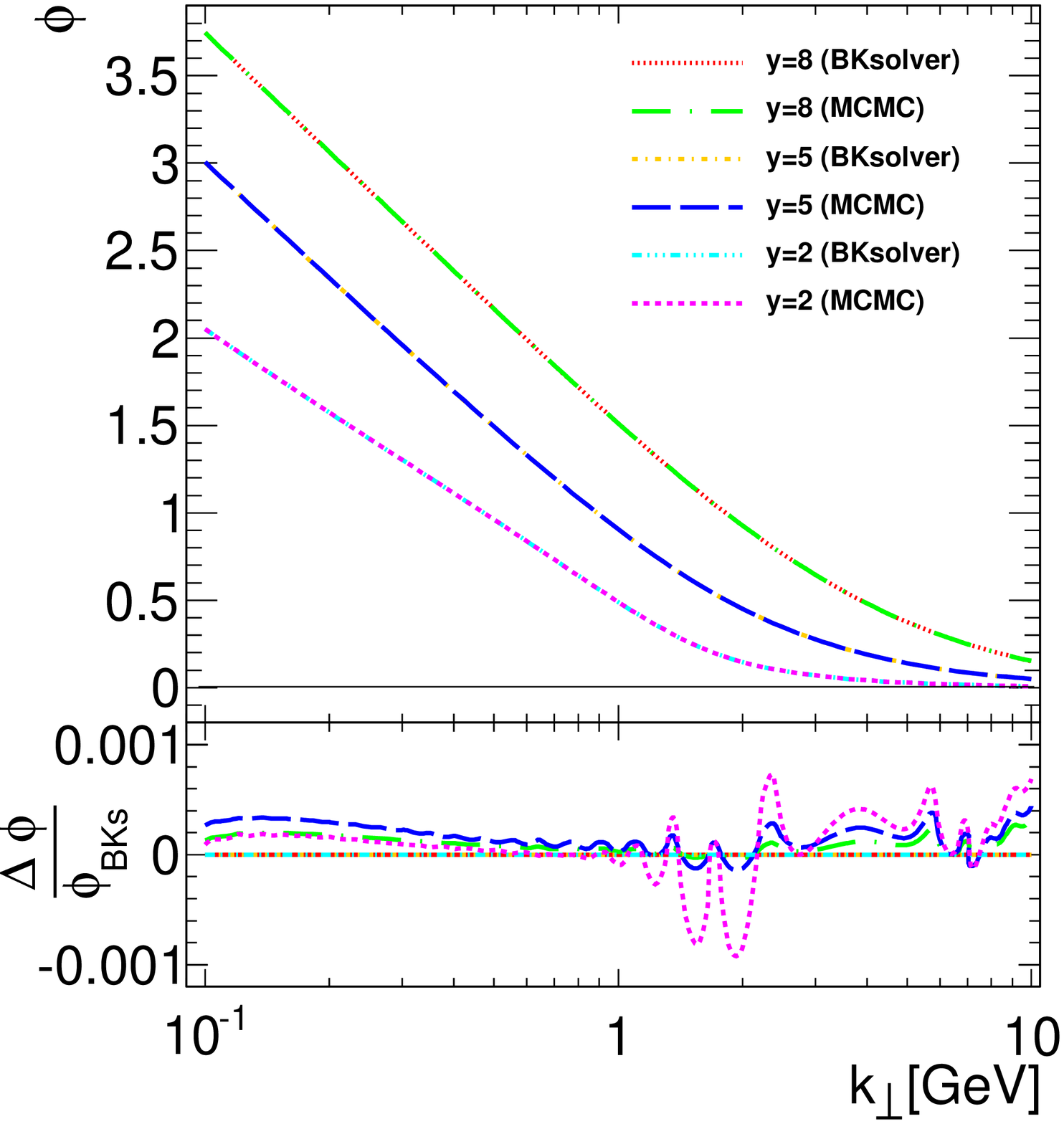}}
\subfigure[]{\includegraphics[width=0.49\textwidth]{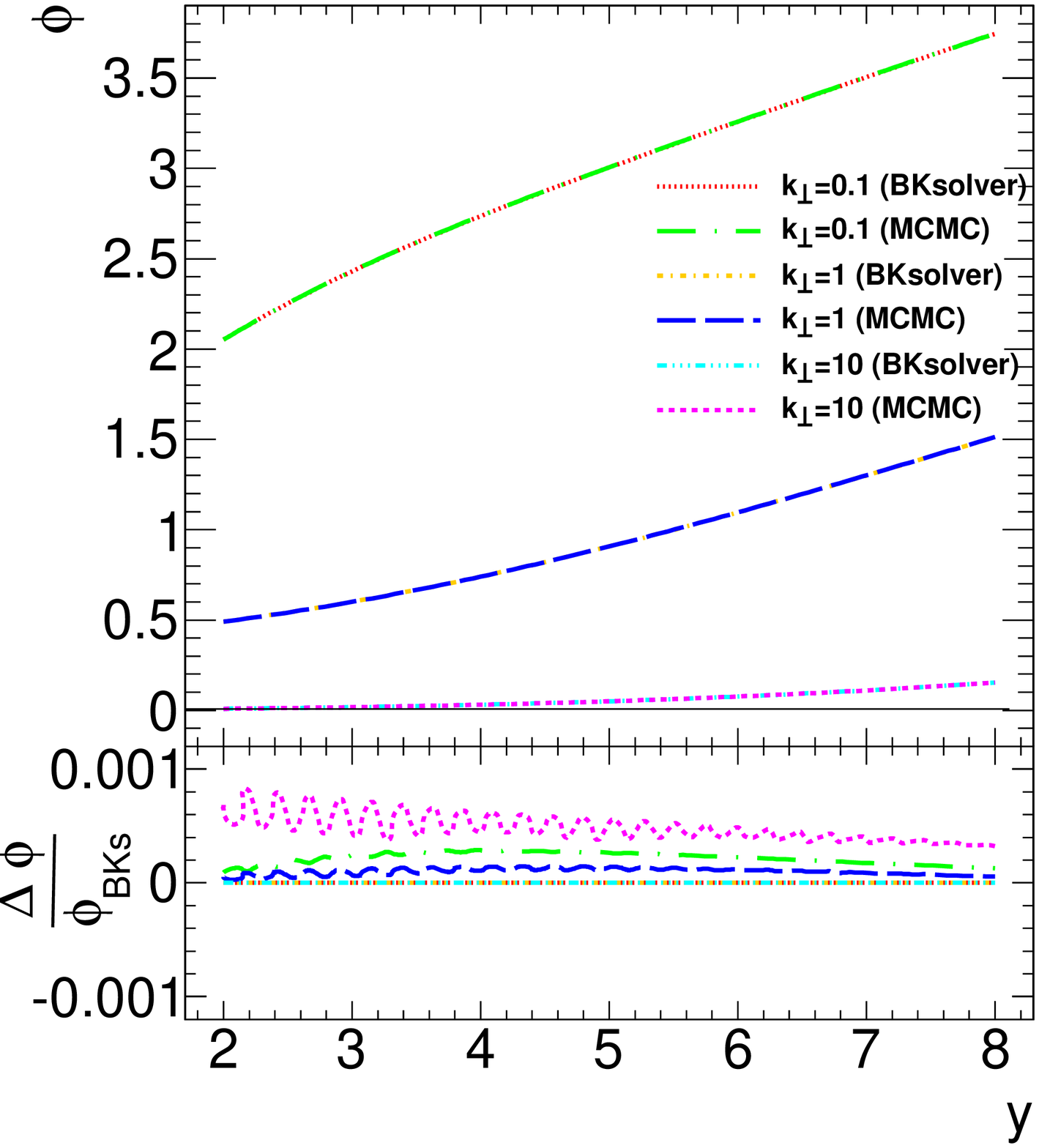}}
\caption{Comparisons of the solutions of the BK equation between the MCMC method and {\sf BKsolver}.} 
\label{MC} 
\end{figure}

In Fig.~\ref{MC} we  show the results of the numerical solution of the BK equation. One can see the profile plots of our MCMC solutions together with the reference solutions obtained with {\sf BKsolver}. 
The latter program evolves the solution of the BK equation in rapidity based on the differential version of the equation. We have run  
{\sf BKsolver} with the same parameter ranges as stated above and required the same number of points in the output lattice. 
The plots presented here have been then obtained using two-dimensional bilinear interpolation. 

In Fig.~\ref{MC}(a) we present the solutions in the $k_{\perp}$-profile for three different rapidity values: $y=2,5,8$. For each of them 
the results from our MCMC algorithm  and from {\sf BKsolver} are shown. In the lower part of the plot one can see the relative difference between the MCMC solution and the reference {\sf BKsolver} one. 
As one can see, these two solutions agree at the level below $0.1\,\%$. 
Similarly, Fig.~\ref{MC}(b) contains three $y$-profiles, each for different $k_{\perp}$ values: $k_{\perp}=0.1,1,10\,$GeV. The results from both the MCMC implementation and {\sf BKsolver} are shown as well as their relative difference. The agreement between the two solutions is again below $0.1\,\%$. 

One might have realized that the results in Fig.~\ref{MC} are shown in the narrower rapidity and $k_{\perp}$ ranges than given at the beginning of this section. We have simply skipped some points on the lattice boundaries where the agreement between the two methods is slightly worse. It is caused by such factors as the interpolation errors and/or the finite number of the lattice points, rather than by a problem of the MCMC algorithm itself. Small fluctuations of the relative differences in both plots (a) and (b)  are due to finite numbers of points in the lattices and approximations of the interpolation procedure. 

Generally, with these results we have proved that the MCMC algorithm is applicable for solving the BK equation 
and, indeed, it gives good numerical results.

\section{Summary and outlook}

In this paper we have developed a general method to solve the two-dimensional non-linear integral equation via Monte Carlo techniques. Our method relies on combining the robust Newton--Kantorovich procedure for solving the non-linear integral equations with the Markov Chain Monte Carlo algorithm. The method is powerful and can be applied to solving complicated, high-dimensional non-linear integral equations, where the traditional methods become inefficient.
It can also open a window to construction of a Monte Carlo event generator based on the non-linear integral equations, 
which will allow to study saturation effects in the fully exclusive processes.

We have applied the MCMC algorithm to the BK equation and compared the results with the ones obtained by using the traditional methods, i.e.\ the solution of the BK equation as an integro-differential equation, implemented in the {\sf BKsolver} package. 
The agreement within $0.1\%$ have been found.
The presented MCMC algorithm is general, and thus it can also be applied to the exclusive form of the BK \cite{Kutak:2011fu} and KGBJS \cite{Kutak:2011fu} evolution equations. This we leave, however, for the future studies.

\section*{Acknowledgments}

We would like to thank Dawid Toton for useful discussions. This research has been partially supported by 
Fundacja Nauki Polskiej (FNP) with the grant
HOMING PLUS/2012-2/6: ``Matrix Elements and Exclusive Parton Densities for Large Hadron Collider"
and by the Polish National Science Centre grant DEC-2012/04/M/ST2/00240.

\end{document}